D.V. Vlasov, L.A. Apresyan, V.I. Krystob, T.V. Vlasova

# Conductivity states changes in plasticized PVC films near breakdown threshold voltages values.

The near threshold "soft breakdown" measurements of PVC films conductivity are investigated. In a wide range of external electric field strength for various rather thick (>20 mkm) PVC films the resistance shows strong nonlinearity and seems to enter high conductive state close to the breakdown threshold. For our "thick" films boundary conditions electrode surface specifics should not be so important as in thin polymer films experiments. Both fast, instant mechanisms of nonlinearity, and effects of accumulation and delay responce were observed. The phenomena corresponding to reversible transitions in a state with rather high conductivity, seems to be similar to ones registered earlier in thin layers of some broad-bandgap polymers. In a range of relatively low field intensity, far from a threshold breakdown, as a result of reversible switches between normal and high resistivity states a polymer film in a standard measuring cell formed a relaxation generator giving a loud enough sound signal with frequency increasing with the increase of external field.

**1. Введение**

В настоящее время явления перехода пленок широкозонных полимеров в состояние высокой проводимости (СВП) под действием различных факторов (одноосного сжатия, деформаций сдвига , сильных и слабых электрических и магнитных полей и т.д., см., напр., [1], а также недавние обзоры [2,3]) можно считать достоверно установленным. Несмотря на то, что физические механизмы собственно проводимости и, в особенности, модели обратимого перехода в СВП в широкозонных полимерах активно обсуждаются [2], достоверного и однозначного описания совокупности аномалий проводимости полимеров до настоящего времени не получено. Изучение спонтанных и стимулированных переходов в СВП в пленках широкозонных полимеров (в частности ПВХ), которые, считаются лучшими и наиболее широко распространенными электроизоляторами, представляется важным как с точки зрения понимания специфики и физических механизмов переключений состояний электропроводности и собственно пробоя, так и для перспектив использования пластифицированного ПВХ в качестве изолятора или универсального антистатического материала. В работах с использованием тонких (микрон и менее) полимерных пленок параметры проводимости и переход в СВП состояние, как отмечается в нескольких работах [2,3] радикально зависит от состояния поверхности электродов ( шероховатость, наличие окислов и т.д.). В отличие от большинства цитированных выше работ, в данной работе выполнены исследования с более «толстыми» пленками 20-100 mkm широкозонных полимеров. При этом очевидно, результаты должны гораздо слабее зависеть от параметров поверхностной структуры электродов и, таким образом, давать больше информации о собственно процессах внутри полимерной пленки. Кроме того, исследования относительно «толстых» пленок имеет непосредственное отношение к практически важным характеристикам надежности полимерной (как правило ПВХ) изоляции промышленной и бытовой электропроводки.

В предыдущей работе авторов [4] исследовались переходы в СВП, происходящие в ПВХ пленках при напряжениях, меньших 1 V/mkm, что значительно ниже известных из литературы [1,2] данных о пороге пробоя для ПВХ. При этом наблюдались как спонтанные, так и управляемые импульсами внешнего напряжения переходы между обычным состоянием и СВП. Специфика исследования электрических свойств тонких полимерных пленок в этой области связана с использованием относительно большого балластного сопротивления - при этом переход в СВП носит полностью обратимый характер, не приводит к разрушению полимерной пленки и может наблюдаться многократно при практически одинаковых пороговых значениях напряженности электрического поля.

**2. Экспериментальная установка и приготовление образцов**

Пленки ПВХ изготавливались как со стандартными фталатными и фосфатными (для сравнения) пластификаторами, так и с использованием нового пластификатора типа А [5], методом полива на стеклянные плоские подложки из 4% раствора ПВХ с пластификатором в тетрагидрофуране (ТГФ). Соотношение ПВХ и пластификатора составляло от 35 до 100 массовых частей на 100 м.ч.ПВХ, соответственно. В измерениях проводимости образцов использовался автоматизированный приборный комплекс описанный в [4,6], позволяющий вести компьютерную запись измерений тока через образец и приложенного напряжения и одновременно визуально контролировать измерения на светодиодных дисплеях. Полимерные пленки-образцы вставлялись в стандартную кольцевую ячейку от ГОСТированного прибора Е6-13 с полной заменой измерительной части на автоматизированный измерительный комплекс.

Программно перестраиваемый стабилизированный источник напряжения в диапазоне 0-2500 V позволял с требуемой скоростью сканировать напряжение, подаваемое на образец и последовательно включенное балластное сопротивление, и одновременно измерять ток, протекающий через образец и балластное сопротивление. В нормальном состоянии образцов все подаваемое напряжение, фактически, приложено к полимерной пленке (сопротивление которого единицы –десятки GOhm). При переходе в СВП ситуация обратная – сопротивление образца может быть менее kOhm [6] и ток ограничивается балластным сопротивлением, на котором падает основная часть напряжения источника.

Для исследования временных характеристик, стабильности и воспроизводимости результатов измерений использовался цифровой осциллограф PS SCOPE фирмы Valleman.

**3. Экспериментальные результаты**

В отличие от измерений электропроводности при низких значениях напряженности поля [4] (~<1V/mkm –т.е. существенно ниже порога пробоя) в описываемых экспериментах исследовались предпороговые характеристики проводимости и собственно сам «мягкий» пробой, т.е. пробой в цепи с последовательно включенным балластным сопротивлением (от 470 MOhm до 27MOhm). С учетом всех включенных в цепь сопротивлений ток через образец не превышал ~ 1mkA, причем при таком «мягком пробое» никаких разрушений полимерного образца не наблюдалось. Более того, после возвращения образца в исходное состояние все его электрические параметры и их зависимости от прикладываемого напряжения достаточно хорошо воспроизводились. Обнаружено, что при напряженности поля порядка 5-10 V/mkm в зависимости от типа образца происходило «гарантированное» переключение пластифицированной ПВХ пленки в проводящее состояние, причем, как отмечалось выше, при использовании соответствующего балластного сопротивления разрушений образца не происходит, и он может «самопроизвольно» вернуться в исходное состояние с сохранением всех электрооптических характеристик. Результаты измерений порогов перехода образцов пластифицированных ПВХ пленок разной толщины в СВП приведены в таблице 1. Отметим, что измеренные пороги «мягкого» пробоя слабо коррелируют с толщиной пленки и слабо зависят от типа пластификатора до тех концентраций, при которых возникает заметная проводимость. Так для переключения образцов с модификатором «А»[5] в концентрациях более 35 м.ч. требовалось использовать другое балластное сопротивление ( вместо используемых 27MOhm ), поскольку сопротивление этих образцов изначально было существенно ниже балласта.

Таблица. 1

| № | Тип пластификатора | Массовая часть (м.ч.) пластификатора на 100 м.ч. ПВХ | Толщина пленки. mkm | U порога переключения, V |
|---|---|---|---|---|
| 1 | Модификатор «А»[5] | 35 | 40 | ~ 800 |
| 2 | Модификатор «А»[5] | 80 | 20 | ~700 |
| 3 | Фталатный (ДОФ) | 80 | 30 | ~700 |
| 4 | Фосфатный | 80 | 20 | ~850 |

В ряде случаев возврат из состояния СВП в исходное состояние при снятии внешнего поля не происходит, и образец, как показали наши эксперименты, сутками может находиться в СВП. При этом его вынимали из измерительной ячейки и даже успешно измеряли сопротивление обычным тестером. В этих измерениях нам удалось подтвердить известный результат о высокой неоднородности распределения проводимости, поскольку проводящие каналы можно было определить непосредственно щупами тестера.

Возврат «толстой» пленки ПВХ из СВП оказывается возможным инициировать также путем охлаждения пленки. Кроме того, порог перехода в СВП также существенно зависит от температуры и, в частности, при охлаждении до температуры жидкого азота переход в СВП в 30mkm пленке ПВХ вообще не наблюдался вплоть до напряжений на ней 2.5 kV.

Общая картина зависимости проводимости от приложенного напряжения и других параметров исследуемых образцов достаточно сложна, хотя собственно зависимость сопротивления образца и приложенного к образцу напряжения (которое, как отмечалось выше, может существенно отличаться от напряжения источника за счет балластного сопротивления) на масштабе от 0V до порога «мягкого» пробоя достаточно универсальна.
В качестве примера на Рис.1 приведены зависимости измеренного отношения тока к напряжению и напряжение собственно на 100мкм ПВХ пленке при изменении общего напряжения стабилизированного источника.

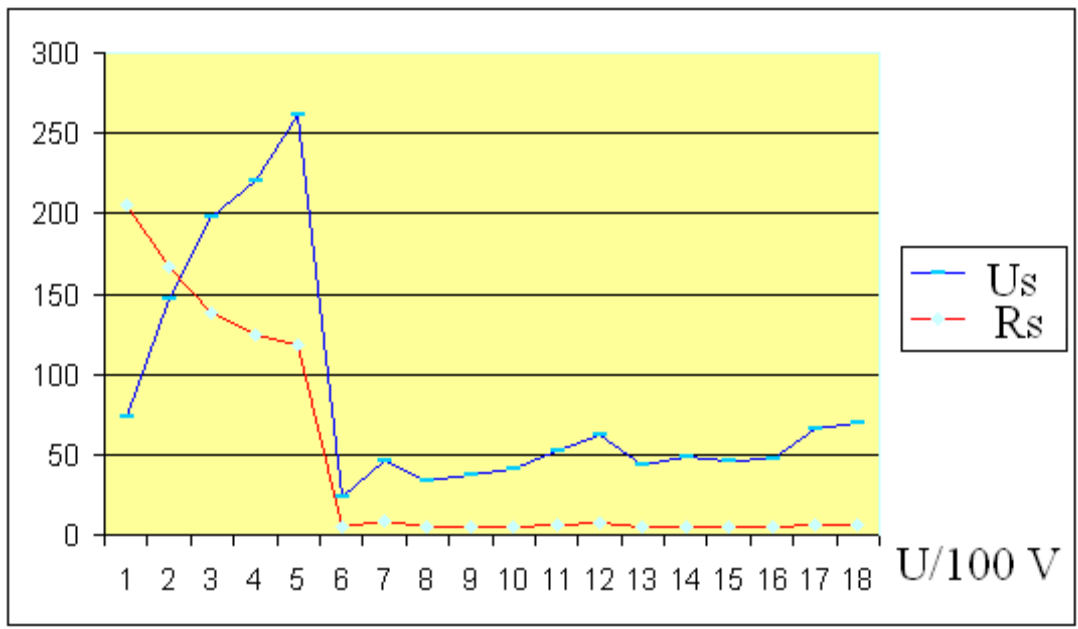

Рис.1. Зависимости сопротивления образца ($R_S$) в произвольных единицах и напряжения на пленке ПВХ $U_s$ V от общего напряжения на ячейке: напряжение -синим цветом, красным вычисленное по измеряемым значениям тока и напряжения значения сопротивления.пленки ПВХ.

На Рис.1 можно выделить три характерных участка общей зависимости параметров образца от напряжения источника :
1. Область плавного уменьшения сопротивления образца (от 0 до точки 5 которая соответствует напряжению источника ~500V). В этой области напряжение на образце по порядку величины «отслеживает» напряжение источника, но эффективное сопротивление образца зависит от приложенного поля и знак этой зависимости может зависеть от различных параметров и от скорости изменения напряжения вследствие хорошо наблюдаемых релаксационных явлений.
2. Область перехода в СВП, или область «мягкого» пробоя, в этой области сопротивление образца становится существенно меньше балластного сопротивления, соответственно напряжение источника практически полностью садится на балласте.
3. Область СВП, в которой могут наблюдаться сильные флуктуации напряжения на образце или «запробойная» область, где при определенных условиях могла возникать устойчивая релаксационная генерация с частотой, зависящей от уровня превышения прикладываемого напряжения над пороговым. Условия возникновения генерации аналогичны условиям получения релаксационного генератора на лавинном триоде (аналогия, развитая в [6]).

Отметим, что в пластифицированных модификатором «А» ПВХ пленках СВП могло сохраняться длительное время *после снятия напряжения* источника. При этом пленка вынималась из измерительной ячейки и даже успешно измеряли локальные значения сопротивление обычным тестером. В этих измерениях нам удалось подтвердить известный результат о высокой неоднородности распределения проводимости и наличие так называемых каналов проводимости, поскольку проводящие зоны на пленке можно было обнаружить непосредственно щупами тестера.

Было обнаружено также, что возврат в нормальное состояние можно ускорить посредством охлаждения образца, причем переход в исходное состояние при охлаждении образца носил опять таки «гарантированный» неотвратимый характер.

С другой стороны, возврат в исходное состояние мог происходить и практически без задержки, т.е. как только напряжение на образце падало до значений близких к нулю, исходные характеристики могли восстанавливаться именно на перепаде напряжения. При этом полимерная пленка в измерительной ячейке выступала как простейший релаксационный генератор – аналог генератора на лавинном транзисторе. В частности, с достаточно большим балластным сопротивлением можно задавать на источнике напряжения как ниже, так и выше пробойного, проходить точку пробоя многократно и воспроизводимо (Рис.1). В зоне выше порога перехода в СВП при соответствующем выборе балластного сопротивления устойчиво наблюдается генерация с циклом – переход в СВП=> падение напряжения на образце практически до нуля=> переход в исходное состояние с высоким сопротивлением=> возрастание напряжения на образце до перехода в проводящее состояние => и снова переход в СВП. Частота генерации определяется постоянной времени разряда емкости образованной электродами измерительной ячейки и сопротивлением образца в СВП и избыточным напряжением над порогом перехода пленки в СВП,- при повышении напряжения – частота релаксационных колебаний увеличивается. Возникающие в таком генераторе релаксационные колебания напряжения приводили к формированию достаточно громкого звукового сигнала вполне различимого на слух. Частота которого увеличивалась с увеличением прикладываемого напряжения. Источник звука в данных экспериментах очевидным образом был локализован в полимерной пленке образца, однако механизм достаточно громкого «звучания» не является тривиальным, поскольку пленка зажата между электродами и совершать колебания по типу мембраны репродуктора или наушника не может.

Для иллюстрации до и над порогового поведения полимерного образца в нескольких циклах программного сканирования напряжения источника от 0 до 2.5 kV на Рис.2 приведены характерные осциллограммы напряжения на образце при нескольких последовательных программных сканах напряжения источника.

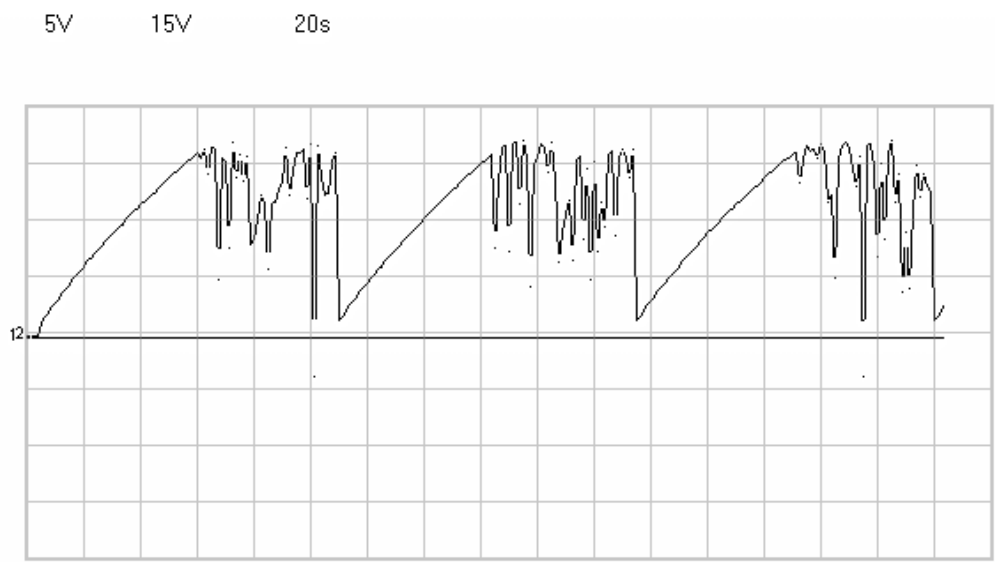

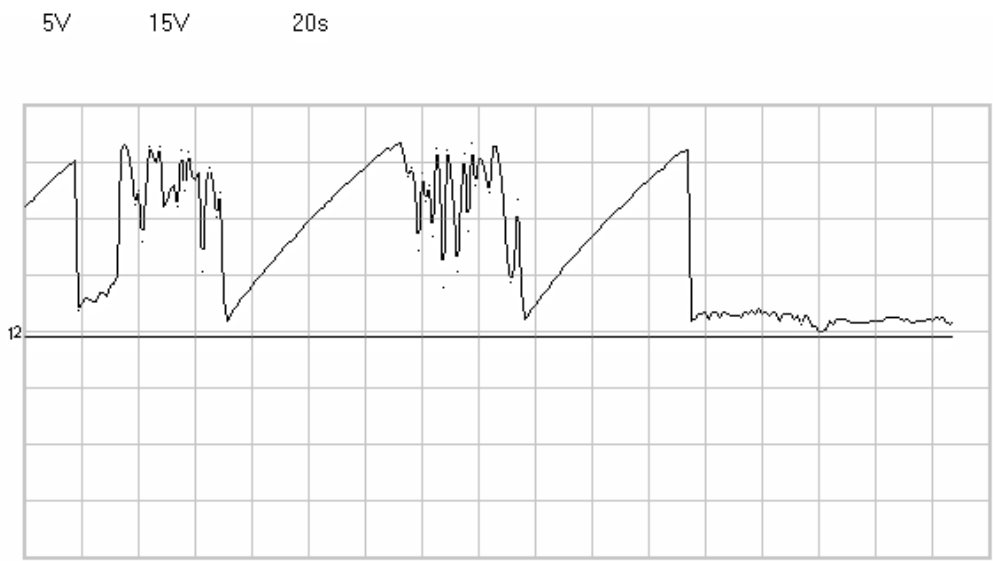

Рис.2 Осциллограмма последовательных циклов сканирования приложенного к пленке ПВХ напряжения.

Выбор временного разрешения связан с желанием отследить несколько циклов программного сканирования напряжения источника. С другой стороны, такое временное разрешение не позволяет отследить по времени циклические колебания напряжения на источнике, которые отчетливо видны при более высоком временном разрешении. Наблюдаемые в процессе генерации перепады напряжения на образце реализуются в диапазоне от 0V до $U_{пор}$, причем амплитуда колебаний далека от синусоидальной и характеризуется случайными сбоями генерации вследствие задержки возврата в непроводящее состояние. В частности, на втором фрагменте осциллограммы видно, что переход в СВП состояние реализуется многократно и обратимо в области «высоких» приложенных напряжений, однако на 6 цикле сканирования наблюдается переход в СВП, время жизни которого существенно больше временных масштабов сканирования.

**4. Обсуждение результатов**

В совокупности измерений параметров исследуемых полимерных пленок установлено, что переход в СВП и обратно носит вероятностный характер. Причем если в [4.6] наблюдались спонтанные переходы в СВП или нам удавалось стимулировать переключение состояния проводимости перепадом напряжения в диапазоне напряжений, заведомо далеких от порога пробоя, то в настоящей работе при повышении напряжения в область порога пробоя переход в СВП был «неизбежен» и вполне однозначен, хотя и небольшой, но отчетливо фиксируемый аппаратурой разброс параметров. Обратный переход в исходное состояние при этом носил сугубо вероятностный характер, и время жизни СВП могло достигать несколько суток. Обнаруженная тенденция сокращения времени жизни СВП в толстых пленках ПВХ при понижении температуры также имеет, судя по проведенным экспериментам, «неотвратимый» характер. Однако при комнатной температуре, как это следует из осциллограммы на Рис.2 , работа ПВХ релаксационного генератора может прерваться на произвольном цикле перехода в СВП с неопределенно продолжительным временем жизни. Тем не менее во всех случаях, необратимых переходов или каких-либо визуальных изменений в полимерных пленках не наблюдалось.

Отметим, что для получения известных из литературы [1,2] трещин, каверн и разрушений при пробое диэлектрика, на наш взгляд необходимо уменьшить балластное сопротивление с тем, чтобы в образце могла выделиться макроскопическая энергия, которая и способна привести к появлению необратимых эффектов в виде трещин, проплавленных областей, деформаций и собственно наблюдаемых каналов прохождения разрядов. Полученные нами результаты аналогичны ситуации с лавинным пробоем электронных приборов типа транзисторов, тиристоров и т.д., где лавинный пробой обратим только при достаточно большом балластном сопротивлении. С этой точки зрения модель электропроводящих свойств полимерной пленки в форме совокупности последовательно включенных p-n переходов, предложенной в [4], получает еще одно подтверждение.

Таким образом показано, что в ПВХ пленках измеряемое сопротивление (т.е. отношение напряжения к протекающему току) демонстрирует сильную нелинейность, причем наблюдается как быстрые, мгновенные механизмы нелинейности, так и эффекты накопления и запаздывания, а также явления, соответствующие обратимым переходам в состояние с высокой проводимостью (СВП). Фактически ГОСТированные измерения сопротивления на базе промышленного прибора Е6-13 при напряжении 100В могут легко давать разброс «в разы», если мы находимся в области медленного изменения U/I, или на несколько порядков, если толщина пленки такова, что поле приближается к порогу пробоя. Спонтанный низковольтный [4] переход в СВП ПВХ пленок, являющихся одним из лучших и наиболее широко распространенных электроизоляторов, представляется достаточно неожиданным и важным как с точки зрения возможных объяснений физических механизмов нелинейности и СВП, так и для использования пластифицированных пленок ПВХ в качестве электроизоляционного или антистатического материала.

В настоящее время строгая физическая картина наблюдаемых явлений отсутствует: этот факт отмечался в недавнем обзоре [2] применительно ко всем широкозонным полимерам. Для объяснения совокупности экспериментальных данных и специфики перехода в СВП вспомним, что пластифицированный ПВХ имеет сложную микро и макро-молекулярную структуру и содержит микроскопические квази-кристаллические и аморфные образования, состоящие, в свою очередь из различных по строению молекулярных объектов (атактической, синдиотактической и изотактической природы). Такое сложное строение позволяет говорить о наличии «квазипроводящих» доменов или глобул с размерами порядка размеров макро- молекулы полимера, на базе которых могут формироваться обсуждавшиеся в литературе каналы электронной прыжковой проводимости. Отметим, что аналогичная модель проводящих глобул в полимерах успешно использовалась в [7] для расчета проводимости полимерных материалов за счет туннелирования электрона через непроводящие барьеры, разделяющие проводящие зоны- глобулы.

В развитие модели квази-проводящих областей- глобул разумно предположить, что пластификатор в случае широкозонных полимеров играет роль, аналогичную роли неоднородного наполнителя, что может приводить к возникновению мелко заглубленных энергетических уровней или ограниченного числа квази-свободных зарядов, которые могут относительно свободно перемещаться в пространственно ограниченных проводящих областях (глобулах) полимера. Длительные (до десятков минут) релаксационные процессы «установления» внутренних полей в полимере, могут свидетельствовать о том, что квази свободные заряды постепенно концентрируются на границах глобул, образуя на соседних глобулах двойные зарядовые слои или «виртуальные» p-n переходы. Таким образом, согласно предлагаемой гипотезе, поведение толстой полимерной пленки во внешнем поле можно рассматривать как набор последовательных p-n переходов, разделенных тонкими (от нанометров до микрон) непроводящими зонами.

В рамках рассматриваемой модели переход в СВП может происходить вследствие приложения локального давления, возникающего в двойном зарядовом слое между двумя глобулами, при этом, как известно из многочисленных экспериментов – повышение давления в тонких пленках полимера однозначно [1-3] приводит к переходу в СВП. Локальные силы притяжения в двойном слое создают гигантское давление и локальный переход в СВП.

Наличие процессов релаксации или запаздывания отклика и низкая подвижность зарядов позволяют объяснить также сохранение СВП при снятии внешнего поля, – перемещенные заряды не успевают вернуться на исходные позиции. Длительное сохранение СВП порядка суток и более возможно в рамках рассматриваемой модели вследствие того, двойной слой продолжает существовать за счет взаимного притяжения зарядов противоположных знаков примыкающих к барьеру глобул. Оценка величины давления, сопутствующее двойным зарядовым слоям, может быть получена, используя аналогию с давлением за счет притяжения пластин конденсатора.

Более того, развиваемая модель локализованных двойных зарядовых слоев позволяет, на наш взгляд, сформулировать гипотезу, объясняющую возбуждение интенсивных звуковых волн в полимерном релаксационном генераторе. В частности, возникновение локализованных зон высокого давления и их синхронная разгрузка при снятии внешнего напряжения в принципе могут объяснить генерацию звука в релаксационном генераторе на плотно зажатой между электродами полимерной пленке. Отметим, что в настоящей работе звуковой релаксационный генератор реализован для толстых полимерных пленок, хотя аналогичные результаты в ряде других работ получены на тонких пленках [1-3]. Неоднородность и случайность локализации зон высокого давления в рассматриваемой модели хорошо увязывается с возникновением случайно расположенных узких проводящих каналов, а также с гистерезисом переходов состояний проводимости. Именно вследствие высокой неоднородности и «случайности» возникновения цепочек глобул с локализованными зарядовыми слоями могут реализоваться спонтанные переходы в СВП, наблюдавшиеся экспериментально в [4,6] в относительно толстых пленках.

Список литературы

1. *Блайт Э.Р., Блур Д*. Электрические свойства полимеров. Физматлит. 2008. 376 с.

2. Лачинов А.Н., Воробьева Н.В.//УФН.2006.Т.176. С.1249-1266

3. А.В.Ванников.//Полимеры с электронной проводимостью и устройства на их основе. Высокомолекулярные соединения.\\ Серия А, 2009, том 51, №4, с 547-571.

4. Власов Д.В., Апресян Л.А., Власова Т.В.,.Крыштоб В.И. Нелинейный отклик и два устойчивых состояния электропроводности в пластифицированных прозрачных поливинилхлоридных пленках, Письма ЖТФ ,2010.(в печати) (см.также ArXiv : 1003.2331, 1003.5482 )

5. *Крыштоб В.И. и др*. Заявка на изобретение №2009131327 от 18.08.2009 г.

6. *D.V. Vlasov, L.A. Apresian, V.I. Krystob, T.V. Vlasova.* Nonlinear response and two stable electrical conductivity levels measured in plasticized PVC thin film samples. ArXiv:1003.5482

7. Ping Sheng.- Fluctuation-induced tunneling conduction in disordered materials.\\ Phys. Rev. B 21, 2180–2195 (1980).